\documentstyle[prl,aps,floats,twocolumn]{revtex}
\begin{document}
\def\be{\begin{equation}}
\def\ee{\end{equation}}
\def\bear{\begin{eqnarray}}
\def\eear{\end{eqnarray}}
\def\E{{\rm e}}
\def\bearst{\begin{eqnarray*}}
\def\eearst{\end{eqnarray*}}
\def\peleven{\parbox{11cm}}
\def\peffec{\peight{\bearst\eearst}\hfill\peleven}
\def\pspace{\peight{\bearst\eearst}\hfill}
\def\ptwelve{\parbox{12cm}}
\def\peight{\parbox{8mm}}
\twocolumn[\hsize\textwidth\columnwidth\hsize\csname@twocolumnfalse\endcsname 
\title
{Phase Transition in a Self--Gravitating Planar Gas}
\author
{Elcio Abdalla$^a$ and 
M. Reza Rahimi Tabar$^b$}
\address
{\it $^a$Instituto de 
F\'\i sica-USP, C.P. 66.318, S\~ao Paulo, Brazil,\\
eabdalla@fma.if.usp.br\\
$^b$ Dept. of Physics , Iran  University of Science and Technology,\\
Narmak, Tehran 16844, Iran.
\\$^b$ Institute for Studies in Theoretical Physics and 
Mathematics
\\ Tehran P.O.Box: 19395-5746, Iran,\\
rahimi@netware2.ipm.ac.ir}

\date{11/05/98}
\maketitle

\begin{abstract}
We consider a gas of Newtonian self-gravitating particles in 
two-dimensional space, finding a phase transition, with a high
temperature homogeneous phase and a low temperature clumped one.
We argue that the system is described in terms of a gas with fractal 
behaviour.

\end{abstract}
\hspace{.2in}
]
\section{Introduction}

The statistical behaviour of systems interacting via classical (Newtonian) 
gravitational forces is rather peculiar as compared with other 
statistical systems, such as neutral gases and plasmas. The central 
distinguishing feature of the former is the fact that there is no 
shielding of the long range gravitational force, while Debye--screening 
rules the long distance behaviour of the electric Coulomb force. In 
lower dimensions such a characteristic of the gravitational force is 
dramatic, due to the rising of the two body potential in Newtonian 
gravity. This leads to conceptual problems, related to the extensive 
nature of energy, as required in statistical mechanics.

In general these statistical problems are not perceptable to exact 
treatment, and only numerical results are known. While classical 
gravitating systems in one space dimension are quite pathological, in two 
space dimensions they become manageable, although the most realistic three 
dimensional case is still beyond reach. In particular, in the latter case 
the phase space volume diverges and one has to use a short distance cutoff.

In two-dimensional space, the thermodynamic functions are analytically 
computable. The thermodynamical properties of a gravitational gas has 
been studied in detail both from a theoretical\cite{her-thi} 
as well as from a numerical \cite{comp,posch,martin,ant-torc} 
point of view. In general, it has been shown \cite{vega,pada} 
that a classical gas of gravitating particles, in the grand canonical 
formalism is related to a field theory described by a Liouville Lagrangian.
In a two-dimensional space Liouville theory is well known \cite{abd} 
and several correlation functions can be exactly 
computed\cite{dorn-otto,david,distler}. 
Indeed, in this particular case the theory is conformally 
invariant, and correlators can be computed in terms of known functions 
of mathematical physics, once properly regularized \cite{dotsenko}. 

At low temperatures the gravitational force is strong, that is, the 
potential energy is big as compared to the kinetic energy, giving rise 
to a collapsing phase, identified by the presence of a single cluster of 
particles floating in a diluted homogeneous background. At high energy a 
homogeneous phase is recovered and the cluster disappears. Using the 
microcanonical ensemble one can show that in the transition region
the system is characterized by a negative specific heat 
(the corresponding instability is of extreme relevance for astrophysics,
\cite{lynden}). 

Hertel and Thirring \cite{her-thi} 
have shown that the canonical and microcanonical 
ensembles are not equivalent in the proximity of the transition.
This thermodynamic inconsistency has been solved in \cite{her-thi}, and 
these results have been successfully confirmed by numerical 
investigations on self gravitating non-singular systems with short range 
interactions \cite{comp,posch}. 

Recently, using numerical calculations, a long range attractive potential 
has been considered, as constructed taking the first few terms of the
Fourier expansion of the logarithmic potential. As a result of the 
simulations, the system turns out to exhibit a transition from a collapsed 
phase at low temperature, to a homogeneous phase \cite{ant-torc}. 

Here we consider a two-dimensional gas of classically interacting 
particles via the Newtonian logarithmic potential, computing its 
thermodynamical properties. Using results known in the framework of 
two-dimensional conformally invariant euclidian field theory, the complex 
integrals may be computed in closed form, and we find a sofisticated structure
of poles and zeros in terms of the temperature. Furthermore, we introduce an 
order parameter, corresponding to the expectation value of the square of the 
two-body distance. We find that at least at one of the singularities of the 
partition function the average distance vanishes, signalizing a clumped 
phase. We are thus able to investigate the collapse of the gravitating 
system. The phase transition point is given in terms of the mass of the 
particles ($m$) and the gravitational constant $G$ by $T_c=\frac 1 
{4k}NGm^2$. We find the exponents of the phase transition.

We furthermore discuss the formulation of gravity in terms of the 
grand-canonical partition function and the related Liouville theory as 
proposed by de Vega et al \cite{vega}, outlining the consequent relations. We 
find, using the same type of complex integrals, a critical temperature in 
this formulation.

\section{Canonical Partition Function}

We consider a gas of nonrelativistic particles with mass $m$ interacting
through Newtonian gravity at temperature $T$. Here, the number of 
particles is fixed and we shall work with the canonical ensemble.
The canonical partition function of the system can be written as
\be
Z= \int [d^2 r_i] [ d^2 p_i] e^{-\beta H_N} \label{can-part-hn} 
\ee
where $ [d^2 r_i] = \Pi_{i=1} ^{N} d^2 r_i $, and the $N$-particle
Hamiltonian $H_N$ is obtained adding the Newtonian potential to the usual
kinetic term, that is,
\be
H_N = \sum_{i=1} ^{N} \frac {p_i ^2}{2m} + 
\frac 12 \sum_{i\neq j} G m^2 \log |r_i - r_j| \label{hamil-n-2d} 
\ee
The potential term must be reqularized, since in general there are 
divergences in either the infrared or in the ultraviolet domain. This is 
done implicitely definning the two-dimensional integrals in the complex 
plane, as done in reference \cite{dotsenko} (see also \cite{abd}).
We thus consider the two-dimensional variables as one single complex 
variable in the complex plane, denoting the procedure by the use of latin 
letters from the end of the alphabet, i.e., $d^2 r_i \rightarrow 
d^2 w_i $. As it turns out, there is an underlining conformal invariance 
in the present problem, easily seen using the complex variables, where is 
is characterized by a $SL(2,C)$ invariance of the Hamiltonian. 
Such an invariance is valid in the infinite volume limit. For finite
volume one should take into account boundary effects, and in that case it would
not be possible to treat the problem exactly. Fortunately, due to the fact
that gravity on a plane constitutes a long range attractive force, the
system is naturally bound to a region defined in terms of the typical
bound state lenght. We thus expect the procedure to be exact for not too
large values of the temperature. Therefore, well above the critical point
departure might be expected from our results. Nevertheless, the critical
temperature is expected to be correct on the above grounds.

Using that invariance, we set $ w_{N+1} =0, w_{N+2}=1$ and $ w_{N+3}=\infty$ 
(whose integration would lead to an infinite overall factor in the partition 
function which can be discarded in the computation of physical 
quantities) and find
\be
Z= (2 \pi m )^{N+3} \int d^2 w_i |w_i|^{ 2 \alpha} |1-w_i|^{2 \beta} \Pi_{i<j} 
|w_i - w_j|^{4 \rho} \label{conformal-partition-before-int} 
\ee
where $ \alpha = \beta =  - \frac {Gm^2}{2 k T} $ and $2\rho=  \alpha$. 
We keep the use of the $\alpha$, $\beta$ and $\gamma$ parameters in order 
to facilitate comparison with the corresponding integrals in the 
literature \cite{abd,dotsenko}.
Using such well known formulae, the partition function $Z$ can be written in 
terms of $\rho $ as 
\bear
Z &\simeq & \Gamma(N+1) [\Delta (1-\rho)]^N \prod_1^N \Delta (i \rho) 
\times\nonumber \\
&&\prod_0^{N-1} [\Delta (1+(i+2)\rho)]^2
\Delta (-1 - (N+3+i) \rho) \label{conformal-partition-after-int} 
\eear
where $\Delta (x) = \frac{\Gamma(x)}{\Gamma(1-x)}$. This function has 
poles at the points $x= -n = 0, -1, -2, \cdots $ and zeros for $x =n = 1, 
2, \cdots$. This is an exact result, as shown in e.g. \cite{dotsenko,abd}. 
However, in order to obtain the singularity
structure (poles and zeros) in further related integrals, it is 
worthwhile to classify them. They arise from the following kind of rather 
naive argumentation.

If one of $w_i$ (say $w$) $\rightarrow 0$ (or $w \rightarrow 1$),
it can be shown that the behaviour of $Z$ is
\be
Z \simeq \pi 
\Delta( 1+\alpha)  \Delta( 1+\beta) \Delta(- 1 - \alpha - \beta ) 
\label{behaviour-z} 
\ee
namely, for $\alpha$ (or $ \beta$) non negative integer a 
zero appears, while for $\alpha$, $\beta$ negative integers, we have poles. 
The poles/zeros in $\alpha$ and $\beta$ are connected with behaviour 
at $w \rightarrow \infty$. We thus obtain some of the functions $\Delta$
given in (\ref{conformal-partition-after-int}). 
If we consider the neighbourhood of some of the $w_i's$ around 0 or 1,
$w_i\sim 0,1$, we  find that $Z$ behaves as
\be
Z \sim  \prod_i \frac {\Gamma(1+\alpha + i \rho)}{
\Gamma(-\alpha- i \rho)}\prod_i \frac {\Gamma(1+\beta + i \rho)}{
\Gamma(- \beta - i \rho)}\quad , \label{nws-to-zero} 
\ee
where $i= 1,2,\cdots, N-1$.

Using the invariance of $Z$ under $w\rightarrow 1/w$, 
one can also show that $Z$ has the symmetry
\be
Z(\alpha, \beta, \rho)= Z( -\alpha -\beta-2 -  \rho(N-1), \beta, \rho) 
\ee

As is evident in the canonical partition function 
(\ref{conformal-partition-after-int}) 
the zeros and poles given in terms of $\rho$ 
depend on the temperature. Those zeros and poles are, in general, 
related to phase transitions. It should be noted that at such points the 
free--energy becomes singular. The simplest possible
phase transition in this system is the collapsing of particles into clumps.
In order to investigate the collapsed phase, it is natural to introduce an 
order parameter describing the average distance of the particles with 
respect to one another. We therefore consider the parameter
$\sum_{i,j} ^N \langle r_{ij} ^2 \rangle$, which allows investigation of the 
collapse of the system. As it turns out, $<r^2>$ has several zeros, 
one of them coinciding with a singularity of the partition function. We 
also show that a class of singularites of $Z$ may be related to the zeroes 
of higher moments of $<r^2>$ i.e $<r^{2q}>$.
Using the invariance of the Hamitonian under translation, we set the 
coordinate of one of the particles to zero and we can write the 
mean square distance as
\be
<r^2>= Z^{-1} <r^2>_0 \label{r2-r20} 
\ee
where
\bear
<r^2>_0 &=&  \int d^2 z \prod_{i=1} ^{N-1} d^2 w_i |w_i|^{ 2 \alpha} 
|1-w_i|^{2 \beta} \nonumber \\ 
&& \prod_{i \neq j} ^{N-1} |w_i - w_j|^{4 \rho}
|z|^{2 + 2 \alpha } |1-z|^{2 \beta} |z - w_i|^{4 \rho}  
\label{r2} 
\eear

Similarly to the case of the  partition function, the integral displayed in
(\ref{r2}) has a complex structure of poles and zeros. 
Let us classify such a singularity structure, using now the hints we have 
gotten based on the computation of the partition function, 
(\ref{conformal-partition-after-int}).

Supposing $w \rightarrow 0 $ (or $w \rightarrow 1 , \infty$) we find
\bear
<r^2>_0 &\sim & 
\prod_{i=1} ^{N-1}   \Delta( 1 + \alpha + i \rho)  
\Delta( 1+\beta+ i\rho ) \times\nonumber\\ 
&&\Delta(- 1 - \alpha - \beta - (N-2+i)\rho)  
\eear
At $w_N=z \rightarrow 0$, $<r^2>_0$ behaves as 
$\frac {\Gamma(\alpha+2)} {\Gamma (-1 - \alpha)}$.
For $z,w \rightarrow 0$, it behaves as
$\frac {\Gamma(1+\alpha+1/2+\rho)} {\Gamma (1 - (\alpha+1/2+\rho))}$.
It is easy to show that analysing the case where several points tend to zero
we see that $<r^2>_0$ must behave as
$\frac {\Gamma(1+\alpha+\frac {1}{i+1}+i \rho)} {\Gamma (1 - 
(\alpha+\frac {1}{i+1}+i\rho))}$,
where $i= 0,1, \cdots , N-1$. The behaviour of $<r^2>_0$ at $1$ similarly
obtained. 

Therefore, for the singularity structure of $\langle r^2\rangle$ 
we arrive at the result
\bear
\langle r^2\rangle_0 &\simeq &
\prod_{i=0} ^{N-2} \Delta (1+\alpha + i \rho ) \nonumber\\
&&\Delta (1+\beta + i \rho )\Delta (-1 -\alpha - \beta - ( N+i-1)\rho ) 
\nonumber \\ 
&& \prod_{i=0} ^{N-1}  
\Delta (1+\alpha + i \rho  + \frac {1}{i+1})\nonumber\\
&&\Delta (-1 - \alpha - \beta - (N-1+i) \rho  - \frac {1}{N-1-i}) 
\eear    
while the normalized $\langle r^2\rangle$ has the form
\bear
&&\langle r^2\rangle \sim  \prod_{i=0} ^{N-1} 
\Delta (1+ (i + 2)\rho + \frac {1}{i+1})\nonumber \\ 
&&\frac  {\Delta (-1 - (N+3+i) \rho  - \frac {1}{N-i-1})}
{\Delta (1+ (N+1) \rho )^2\Delta (-1- 2(N+1) \rho )} 
\eear
for $i=0$ it has a zero at $1+\alpha+1$= positive integer $ = 1,2 \cdots$ 
therefore the possible value of $\rho$ is given by 
$ 2\rho = \alpha = -1, 0, 1, \dots$.

The expression for $r^2$ has several zeros. They can certainly not be all
physically relevant, since most are just consequences of the
regularization of the integrals (\ref{r2}). After encountering a zero, we
loose the physical relevance of the integral. Below the critical
point, the theory is in a clumped phase, while above the particles are
far apart. Therefore, there should be no further critical point. This
fixes our critical point to be, at large $N$
\be 
kT_c =\frac 14 N Gm^2\quad .\label{trans-temp} 
\ee    
For large $N$ this coincides with an old result of Salzberg\cite{salz}, 
from wich a possible phase transition point can be obtained.

Note that $\langle r^2\rangle$ behaves as $ \vert T-T_c\vert $ in terms of 
the temperature.

The higher moments of $<r^2> $ have the singularity  structure
\bear
\langle r^{2q}\rangle & \sim & \Pi_{i=0} ^{N-1} \frac {
\Delta (1+ (i + 2)\rho + \frac {q}{i+1})}
{\Delta (1+ (N+1) \rho )} \nonumber \\ &&
\frac {\Delta (-1 + (N+3+i) \rho  - \frac {q}{N-i-1})}
{\Delta (-1- 2(N+1) \rho )} 
\eear
There are zeros at the values $\rho = -\frac12 q, -\frac 12 q+1,\cdots $. 
If $q$ is a multiple of $(i+1)$, for $i= 1,\cdots N-1$ there are further 
singularities. Depending on the exact value of $q$ the behaviour in
terms of the departure from the critical temperature changes, which shows 
that the system has multi-fractal nature.
The zeros of $<r^{2q}> $ are coincident with a class of singularities of $Z$.
For $q=2(N-1)$ the behaviour of $\langle r^{2q}\rangle$ is not linear in 
$\vert T-T_c\vert$.

\section{ The Grand Partition Function}

The grand partition function of the system can be written as
\be
Z_G = \sum _{N=0} ^ {\infty } \frac {z^N}{N!} \int [d^2 r_i] [ d^2 p_i]  
e^{-\beta H_N} \label{grand-partition} 
\ee
where $H_N$ is given by (\ref{hamil-n-2d}) 
and $z = e^\mu$ is the fugacity.

Acoording to \cite{vega,pada} this many body problem can be transformed
into a field theoretic one. Using the definition of density 
$ \rho( \bf r) = \sum_{i=1} ^N \delta ( \bf r - \bf q_i) $, it has been
shown that $Z_G$ can be written in terms of Liouville field theory,
\be
Z_G= \int D{\phi'} e^{-\frac {1}{T_{eff}} \int \lbrack\frac{1}{2} 
\left(\nabla \phi'\right)^2 - {\mu'} ^2 e^{\phi'}\rbrack}
\label{liou-partition} 
\ee
where $T_{eff}= 2 \pi \frac{G m^2} {k T}$ and $\mu'^2 = z G m^3$.
we rescale $\phi'$ as $\phi'= \sqrt{\frac{T_{eff}}{4 \pi}}\phi$. 
Therefore, the action can be written as
\be
S_L = \frac {1}{8 \pi} \int d^2 x ( |\nabla \phi|^2 + \mu e^{b \phi}) 
\label{liou-action} 
\ee
where $b= \sqrt{\frac {T_{eff}}{4 \pi}}$ and $\mu = \frac {- 8 \pi 
\mu'^2}{T_{eff}}$. In \cite{vega}, it has also been shown that the 
correlations of the density can be written in terms of vertex operators 
of the Liouville theory.

It is conventional to add term $\frac{Q}{4\pi} R \sqrt{g} \varphi$
to the Liouville Lagrangian density, where $R$ is the scalar curvature 
of background metric $g_{\mu\nu}$, and the parameter $Q$ adjusted
to ensure that all physical quantities be independent of a particular choice
of the background. However it is possible to choose a specific background
which is flat everywhere except for few selected points \cite{zamozamo}.

The Liouville field $\phi(z,\bar z)$ is a logarithmic operator 
\cite{khorra} and 
varies under holomorphic coordinate transformation $z \rightarrow w(z)$ as
$\phi(w,\bar w) =  \phi(z,\bar z) - \frac {Q}{2} \log |\frac {dw}{dz}|^2$
where $ Q= b + \frac {1}{b}$. $Q$ parametrizes the central charge of the 
theory by the well knwn relation $c=1 + 6 Q^2$.

The introduction of $Q$ in the Liouville action is done for taking into 
account the fact that the
theory corresponds to a Coulomb gas where there is no zero charge sector,
except for the vacuum. It corresponds to different boundary conditions, and
takes account of the zero modes of the theory \cite{abd}. In the description
of ref. \cite{vega} there is no zero mode in the inverse propagator, fixed as
being the classical gravity potential, which naturally matches the fact that
the background curvature has to vanish in Minkovski space. 

We suppose that the theory can be described as a compactified Euclidian
space, in which case the $Q$-term can be seen as a requirement of
renormalization. The question to be answered is whether the large
compactification radius limit is smooth or not. Since our argumentation
are based on local Green functions, as in (\ref{deltansquared}) 
below, we do not expect any major difference to occur. On the other hand, the 
argumentation based on the conformal dimensions (as in (\ref{rho-to-q})) 
below and following equations) may depend on the above limit. However, 
it is rewarding to see that the
results are in accordance with general expectations, leading to the
conjecture that the large compactification radius limit is smooth.

It is nevertheless necessary to stress that the introduction of the
$Q$ term is non trivial, and might lead to a change of the problem. The
aim here is to describe the results obtained comparing them with known 
results. We know in fact from \cite{vega} that Liouville theory describes 
gravity, and any effort in the direction of understanding the model is worth
undertaking.

It is well known that the exponential Liouville operators
$V_\alpha(x)= e^{2\alpha \varphi(x)}$
are the spinless primary conformal fields of dimension
$\Delta_{\alpha} = \alpha (Q-\alpha)$.
The two, three and four-point correlation functions of Liouville field
theory for given $\alpha$ have been calculated  in \cite{dorn-otto,zamozamo}.  

In addition, we find that the dependence on 
the scale $\mu $ of any correlation function in
Liouville theory \cite{kpz,distler} is
\be
\langle \Pi_{i=1} ^N e^{2 \alpha_i  \varphi (x_i) }\rangle_Q\sim (\pi \mu)^
{(Q-\Sigma_{i=1} ^{N}\alpha_i)/b}\quad . 
\ee

We note that the Liouville theory is self-dual under $b \rightarrow {1}/{b}$. 
Indeed, we consider the partition function of the Liouville theory.
According to \cite{dorn-otto} the partition function has the form
\be
Z=-\frac{\mu}{\sqrt {2} \pi^2 (b+1/b)} \big {(}\frac {\pi \mu \Gamma (b^2)}
{\Gamma(1-b^2)}\big{)}^{1/b^2} \frac {\Gamma (-1/b^2)}{\Gamma(1/b^2 -1)} 
\label{partition-liou-in-b} 
\ee

Under transformation of $b $ and $\mu $ as given by
\be
b \rightarrow  \frac{1}{b}, \hskip 1cm  \mu \rightarrow \bar {\mu} = 
\frac {1}{\pi \Delta (1/b^2)} (\pi \mu \gamma(b^2))^{1/b^2} 
\ee
it is easy to show that the partition function transforms as
\be
Z(b,\mu) \Rightarrow Z(\frac {1}{b}, \bar \mu)=-\frac {1}{b^2} Z(b,\mu) 
\ee

This duality transformation was first observed by Zamolodchikov\cite{zamozamo}.
We further observe that there exists a sequence of critical  
values for $b$, so that the partition function becomes singular, that is
\be
b_N ^2= \frac {b_c^2}{N} 
\ee
where $b_c=1$.  As in the case of the canonical partition function, the
grand-canonical $Z_G$ has zeros.

In order to understand the nature of phase transition in the grand-canonical 
ensemble we define the variance of the number of particles as a
parameter, where in the clumping point $(\Delta N)^2=(<N^2>-<N>^2)$.
This variance has a peculiar behaviour. It is easy to show 
that the $(\Delta N)^2$ can be witten in terms of 
two--point density correlation functions as:
\be
(\Delta N)^2 \sim \int \int d^2 x_1 d^2 x_2 < \rho(x_1) \rho(x_2)>
\label{deltansquared} 
\ee
Now using the results of \cite{dorn-otto} for two point correlation 
functions of Liouville vertex operators, one can show that at the 
transition point $\langle\rho(x_1) \rho(x_2)\rangle \rightarrow 0$, which 
shows that particles collapse to local clusters.

Now it is possible to investigate the multi--fractal nature 
of 2D-Gravitating gas  considering  moments of the density in a
given scale $R$. Using the conformal dimensions of Liouville operators we 
can show that
\be
\int _R d^2 x < \rho^q> \sim R ^{-\tau(q)} \label{rho-to-q} 
\ee
where $\tau (q) = 2 (1-b^2 q) (q-1) $ which is valid only for 
$q \leq \frac {b^{-2} + 1}{2}$. 

Equation (\ref{rho-to-q}) allows us to determine the distribution 
functions of density $p(\rho)$, such that
\be 
\langle\rho^n\rangle = \int_0 ^{\infty} \rho^n p(\rho) d\rho\quad ,
\ee
and
\be
p(\rho)= f(\rho){\large e^{-\frac{1}{\log R^a} \log^2(\rho R^d)}}
\ee
where $f(\rho)$ is a smooth function of $\log(\rho)$. Moreover $a=16b^6$, 
$d=2b(1+4b^2)$ (see also \cite{17}).
This is the famous log-normal distribution,
considered as a characteristic feature of a disordered system.
It turns out that for a dilute gas  $\rho\simeq \frac{1}{R^{2b(1+4b^2)}}$,
and we find the Gaussian distribution functions, whose variance is 
controlled by $R$.\\

\vskip .3cm

{\bf Acknowledgements:}
E.A. wishes to acknowledge Conselho Federal de Desenvolvimento 
Cient\'\i fico e Tecnol\'ogico (CNPq-Brazil) for partial financial support 
and Prof. A. Aghamohammadi for the kind hospitality at the Institut for 
Theoretical Physics and Mathematics, Tehran, Iran. 
M.R.R.T. would like to thank B. Davoudi, F .Darabi, J. Davoudi,
R. Ejtehadi, A. Langari, R. Mohayaee and S. Rouhani for useful discussions.


\end{document}